\documentclass[sigplan,screen,margin=1in]{acmart}
\AtBeginDocument{%
  }

\setcopyright{none}
\copyrightyear{2018}
\acmYear{2018}
\acmDOI{XXXXXXX.XXXXXXX}
\acmConference[Conference acronym 'XX]{Make sure to enter the correct
  conference title from your rights confirmation email}{June 03--05,
  2018}{Woodstock, NY}
\acmISBN{978-1-4503-XXXX-X/2018/06}

\renewcommand\footnotetextcopyrightpermission[1]{} 

\usepackage{newtxtext,newtxmath}
\usepackage{graphicx}
\usepackage{amsmath}
\usepackage{xspace}
\usepackage{natbib}
\usepackage{xcolor}

\usepackage{subfig}
\usepackage{tikz}

\usepackage{multirow}

\begin{document}

\title{Estimating The Energy Consumption of Quantum Computing from A Full System Aspect}


\author{Siyuan Niu}
\affiliation{%
  \institution{University of Central Florida}
  \city{Orlando}
  \state{Florida}
  \country{USA}
}
\email{siyuan.niu@ucf.edu}

\author{Di Wu}
\affiliation{%
  \institution{University of Central Florida}
  \city{Orlando}
  \state{Florida}
  \country{USA}
}
\email{di.wu@ucf.edu}

\author{Ozgur Ozan Kilic}
\affiliation{%
  \institution{Brookhaven National Laboratory}
  \city{Upton}
  \state{New York}
  \country{USA}}
\email{okilic@bnl.gov}

\author{Kwangmin Yu}
\affiliation{%
 \institution{Brookhaven National Laboratory}
 \city{Upton}
 \state{New York}
 \country{USA}}
\email{kyu@bnl.gov}

\renewcommand{\shortauthors}{Niu et al.}

\begin{abstract}

Quantum computing promises disruptive capabilities, yet its energy
footprint has received far less attention than its asymptotic
speedups. We present a first-order, full-system energy model for
quantum computing in an high performance computing (HPC) context. The model separates costs
common to NISQ and FTQC, such as
system maintenance and classical processing, from regime-specific
ones such as error mitigation for NISQ and error correction for
FTQC. We instantiate the model on $96$- and $100$-qubit Heisenberg time-evolution simulations on IBM
Eagle r3 and a representative VQE workload, and sketch the FTQC
energy pipeline. We find that NISQ energy is dominated by the QEM
sampling multiplier, while FTQC cost shifts to physical-qubit
overhead set by the code distance and magic states. Our model
provides actionable insights into the energy consumption of both
NISQ and FTQC workloads, and paves the way toward energy-efficient
quantum advantage.

\end{abstract}


\maketitle
\section{Introduction}
Quantum computing offers disruptive computing capabilities, such as prime factorization and combinatorial optimizations, quantum simulation and cryptography, which are beyond the reach of classical computing.
With the recent improvements on both quantum algorithms and hardware, prototyping utility-scale quantum computers is on the horizon.
Despite tremendous efforts in the computing capability, such as functional correctness and performance, one untouched research question of quantum computing is: \textit{how much energy does quantum computing consume?}
This question is critical to energy, cost, and power grid allocation in high performance computing (HPC) systems, where quantum computers are integrated as important sub-components.
Understanding the energy landscape will answer a pivotal question: \textit{when and where to apply quantum computing}.

However, quantum energy estimation is complicated by the heterogeneity of the environment in which quantum algorithms are executed. Quantum processors are rarely operated in isolation; they increasingly act as accelerators inside HPC-cloud-quantum substrates, in which each subtask is, in principle, routed to its most efficient compute resource. 
The prototype of such substrates have already been demonstrated in industries, such as the quantum-centric supercomputing from IBM~\cite{seelam2026reference} and NVQLink from Nvidia~\cite{caldwell2025platform}. Energy accounting across such substrates is fragmented
across siloed mechanisms: facility power-usage effectiveness (PUE) metrics report infrastructure efficiency without resolving job-level consumption~\cite{avelar2012pue}; node-level counters capture only partial IT loads~\cite{david2010rapl}; cloud billing approximates resource usage rather than energy~\cite{dabbagh2014cloud}; and quantum services seldom expose cryogenic or control overheads~\cite{enriquez2023estimating}. Consequently, no unified cost model exists that simultaneously
accounts for classical data movement, network transfer, and
quantum-specific contributions such as cryogenic cooling and control electronics.

Existing quantum-energy models~\cite{martin2022energy,enriquez2023estimating} address only a subset of these concerns. First, they rely on quantum-volume worst-case estimation, which overstates energy consumption by neglecting recent architectural advances such as dynamic circuits with mid-circuit measurement, feed-forward control, and conditional execution~\cite{niu2024ac, shirgure2026characterizing}, all of which reduce the effective circuit size. Second, they are formulated for a single qubit technology, typically superconducting, and do not generalize to trapped-ion, photonic, or neutral-atom platforms, including maintenance budgets for ultra-high vacuum, cryogenic cooling, and laser stabilization that differ by orders of magnitude. Third, they restrict their scope to the quantum subsystem and omit the classical pre- and post-processing on which every realistic workload depends.

The energy footprint of quantum computation is therefore determined by three interacting factors: the \emph{quantum application}, which fixes gate count, circuit depth, and qubit count; the \emph{quantum hardware}, which fixes per-gate energy and maintenance overhead through the underlying qubit technology; and the \emph{operating regime} under which the device executes. In this work, we focus on
both noisy intermediate-scale quantum (NISQ) computing and fault-tolerant quantum computing (FTQC) regimes.

The energy profiles for NISQ and FTQC are different.  NISQ devices tolerate hardware imperfection and rely on quantum error mitigation (QEM), comprising techniques such as zero-noise extrapolation, Pauli
twirling, dynamical decoupling, and measurement
mitigation~\cite{Temme-error-mitigation, Kandala-error-mitigation, kim2023scalable,chowdhury2024enhancing, choi2025quantum}. QEM
introduces no qubit overhead but inflates circuit and shot counts;
the dominant energy cost therefore arises from additional circuit
execution and repeated sampling. FTQC, in contrast, suppresses noise
through QEC. The associated qubit overhead is substantial depending on the QEC code, and is accompanied by continuous syndrome measurement, real-time classical decoding, and magic-state distillation/cultivation~\cite{gidney2024magic,bravyi2012magic}. In this
regime, energy consumption is dominated by physical-qubit maintenance,
decoder hardware, and the multiplicative overhead of logical
operations. Although both regimes share a common substrate (gate
execution, qubit idling, system maintenance, and classical
processing), they differ in their treatment of error, which gives
rise to distinct optimization strategies and distinct projections as
the technology scales.

The contribution of the paper is listed as follows:
\begin{itemize}
\item We summarize the energy sources of a quantum workload, separating costs common to NISQ and FTQC from
regime-specific overheads.
\item We quantify the NISQ side with two case studies: $96$- and $100$-qubit Heisenberg time-evolution
simulations on IBM Eagle r3 hardware, and a representative VQE workload.
\item We sketch the FTQC energy pipeline and show how the
dominant cost shifts from sampling to physical-qubit overhead.
\end{itemize}

\section{Where Does the Energy Come From?}
\label{sec:sources}
A quantum workload draws energy from several distinct sources. We
separate them into a small \emph{common overhead} that is paid by
both NISQ and FTQC executions, and a \emph{regime-specific
execution} term that captures the algorithm-dependent cost of
running the workload under either QEM
or QEC. 


\subsection{Common Overhead in NISQ and FTQC}
\label{sec:common}

\paragraph{System maintenance.} Different qubit technologies impose drastically
different maintenance budgets:
superconducting qubits require dilution refrigerators that draw
$\sim$10-25\,kW continuously; trapped ions require ultra-high-vacuum
chambers and stabilized lasers; photonic qubits require single-photon
sources and detectors with their own cooling.
This term is technology-dependent and largely independent of the
specific algorithm being run, which makes it a consequential
denominator when comparing platforms~\cite{martin2022energy}.

\paragraph{Classical pre/post-processing.} Every quantum job is
flanked by classical work: circuit transpilation/compilation, parameter
optimization, results aggregation, and integration with HPC pipelines
(file I/O, network transfer, scheduler overhead). For hybrid
HPC-cloud-quantum substrates, this can be expressed as a job-level
energy boundary
\begin{equation}
E_\text{cls} = \mathrm{PUE}(t)\cdot E_\text{IT} + E_\text{shared}
+ E_\text{net,WAN} + E_\text{storage},
\label{eq:Ecls}
\end{equation}
with $\mathrm{PUE}(t)$ the time-varying facility
efficiency~\cite{greengrid2014pue}, $E_\text{IT}$ aggregating per-node
package/DRAM/GPU/NIC energy via RAPL/NVML and equivalent
counters~\cite{david2010rapl,dabbagh2014cloud}, and the remaining terms covering
shared infrastructure, network, and storage.

\subsection{NISQ-Specific Overhead}
\label{sec:nisq-extra}

\paragraph{Gate execution.} Every quantum operation has a non-zero
energy-per-gate $E_g$, dependent on qubit technology. Reported
values include $E_g \approx 0.18$~J for superconducting two-qubit
gates and $E_g \approx 15$~J for trapped-ion two-qubit gates~\cite{enriquez2023estimating}, both inclusive
of system cooling overhead. However, gate-energy estimation remains an
early-stage research area, and reported values vary
substantially across studies and
methodologies~~\cite{jaschke2023quantum,silva2023classical}.


The bare per-circuit physical-gate energy is
\begin{equation}
E_\text{gate} \;=\; \sum_g E_g \cdot N_g,
\end{equation}
summed over all gate types $g$ with their counts $N_g$ in the
executed circuit.
\vspace{-7pt}


\paragraph{Error mitigation.} QEM accepts imperfection of near-term
quantum devices and adopts methods of mitigating or
suppressing quantum noises~\cite{Temme-error-mitigation,endo-error-mitigation}. The cost mainly comes from sampling and circuits. We first isolate the baseline sampling cost that any NISQ workload
incurs, and then quantify the QEM-specific multipliers that act on top
of it. A single quantum circuit returns one bitstring per execution,
so estimating an expectation value $\langle H\rangle$ to statistical
precision $\epsilon$ requires $S \sim \mathcal{O}(1/\epsilon^2)$ shots,
typically $S \in [10^3,10^5]$. This sampling cost is intrinsic to
quantum measurement; it is present even in the absence of any
mitigation, so we treat $S$ as a baseline rather than as a QEM
mechanism. On top of this baseline, we analyze the overhead of two QEM techniques:
\begin{itemize}
\item \textbf{Zero-Noise Extrapolation (ZNE).} The same circuit is
executed at multiple effective noise levels (e.g., scaling factors
$\{1,3,5\}$ via gate folding), giving a multiplicative factor
$\sum_k \alpha_k$ on the raw gate count, where $\alpha_k$ is the
fold of copy $k$.
\item \textbf{Pauli Twirling (PT).} Each ZNE copy is replicated $P$
times with random Pauli insertions, which converts coherent error
into a stochastic Pauli channel and stabilizes the ZNE
extrapolation. This contributes a multiplicative factor $P$ on top of ZNE.
\end{itemize}

Combining the baseline sampling cost with the two QEM multipliers, the NISQ-specific execution energy is therefore
\begin{equation}
E^\text{exec}_\text{NISQ} \;=\; \Big(\sum_k \alpha_k\Big)\cdot P \cdot S \cdot E_\text{gate}
\label{eq:nisq}
\end{equation}

Other error mitigation techniques like dynamical decoupling (DD) and Matrix-free Measurement Mitigation (M3)
do not duplicate the full circuit, so they do not enter
Eq.~\eqref{eq:nisq} as multiplicative factors, but they are not free.
DD inserts pulse sequences into idle windows; this incremental gate
cost is absorbed into $E_\text{gate}$. M3 requires a one-time calibration of the
readout assignment matrix $A$ on the relevant measurement subspace,
consuming $S_\text{cal}$ shots that contribute to $E_\text{gate}$ but
are amortized over subsequent evaluations; the sparse classical
inversion of $A$ is folded into $E_\text{cls}$ in Eq.~\ref{eq:Ecls}.

\subsection{FTQC-Specific Overhead}
\label{sec:ftqc-extra}
FTQC encodes a logical qubit into many physical qubits and detects
errors via continuous syndrome measurement. 
Different QEC code families exhibit different code rates and span across wide code distances.
Despite of the differences, three energy-relevant cost centers result:
\begin{itemize}
\item \textbf{Encoding overhead.} We focus on the most explored surface code, which 
requires $\Theta(d^2)$ physical qubits per logical qubit at distance $d$. Each logical
gate is implemented through a sequence of physical operations that
scales with $d$, multiplying $E_\text{gate}$ accordingly. 
The logical error rate of surface code scales as $p_L \propto (p/p_\text{th})^{(d+1)/2}$.
With a threshold of $p_\text{th}\!\approx\!1\%$, a target logical error rate of $10^{-12}$, and a physical error of $10^{-3}$, the surface code needs $d \approx 25$~\cite{fowler2012surface,
gidney2021factor}, implying roughly $10^3$ physical qubits per logical qubit.
The encoding cost will be lower for qLDPC codes. The unit of a logical circuit is one spacetime cell whose energy is denoted by $E_\text{cyc}$ and the total spacetime volume of the compiled program is $V_\text{ls}$.
\item \textbf{Real-time decoding.} Classical decoders process syndrome information and produce corrections inside a strict latency budget which is on the order of $\mu s$ for superconducting
devices~\cite{google2025quantum}. 
Further considering the syndrome measurement to avoid backlog problem~\cite{terhal2015quantum}, decoding has to complete within 400$ns$ for superconducting-based quantum computers~\cite{ghosh2012surface}.
Moreover, decoding cost grows with code distance and varies
with implementation (CPU, FPGA, ASIC) and physical placement (cold,
4\,K stage, or room-temperature)~\cite{enriquez2023estimating}. 

\item \textbf{Magic-state distillation/cultivation.} Non-Clifford operations (e.g., $T$-gates) are not transversal in most codes and must be realized using prepared high-fidelity magic states. Conventional distillation factories consume large numbers of physical qubits and many code cycles per distilled state~\cite{litinski2019magic}; more recent magic-state cultivation protocols grow $T$ states in place and reduce the qubit overhead significantly but remains sampling overhead~\cite{gidney2024magic}. Either approach
contributes a separate $E_\text{ms}$ term whose share grows with $T$-gate count.
\end{itemize}
The FTQC-specific energy is therefore
\begin{equation}
E^\text{exec}_\text{FTQC} \;=\; V_\text{ls}\!\cdot\!E_\text{cyc}
\;+\; N_T\!\cdot\!E_\text{ms}
\;+\; E_\text{dec}
\label{eq:ftqc}
\end{equation}

where $V_\text{ls}$ is the total space time volume of the compiled logical circuit, $E_\text{cyc}$ is the energy of a spacetime cell, $N_T$ is the number of magic states, and $E_\text{dec}$ is the decoding energy.
\vspace{-5pt}
\paragraph{Putting it together.} For a given quantum workload, the total energy is
\begin{equation}
E_\text{tot} \;=\; \underbrace{E_\text{sys} + E_\text{cls}}_\text{common overhead}
\;+\; \underbrace{E^\text{exec}_\text{NISQ}\ \text{or}\ E^\text{exec}_\text{FTQC}}_\text{regime-specific execution}
\label{eq:total}
\end{equation}
with at most one of the two regime-specific terms active. NISQ
workloads spend most of their joules on the QEM multiplier; FTQC
workloads spend joules evenly.

\begin{table*}[t!]
\centering
\begin{tabular}{ c || c | c | c || c | c | c }
\hline
& \multicolumn{3}{c || }{OBC (100 sites)} & \multicolumn{3}{ c }{PBC (96 sites)} \\ [0.3ex] 
\hline
 ZNE fold & ~~~~~~~~1~~~~~~~~ & ~~~~~~~~3~~~~~~~~ & ~~~~~~~~5~~~~~~~~ & ~~~~~~~~1~~~~~~~~ & ~~~~~~~~3~~~~~~~~ & ~~~~~~~~5~~~~~~~~ \\ [0.3ex] 
\hline \hline
No. of gates with PT & 24,040 & 49,300 & 74,560 & 25,400 & 49,880 & 74,360 \\ [0.3ex] 
\hline
Energy (kJ) with PT & 432,720 & 887,400 & 1,342,080 & 457,200 & 897,840 & 1,338,480 \\ [0.3ex] 
\hline
Total Energy (kJ) & \multicolumn{3}{c || }{2,662,200} & \multicolumn{3}{ c }{2,693,520} \\ 
\hline
Power (MW) & \multicolumn{3}{c || }{2.686} & \multicolumn{3}{ c }{2.684} \\ [0.3ex] 
\hline
\end{tabular}
\caption{The total energy consumption of the Hamiltonian simulation. The total QPU time of the simulations is $16$ minutes $44$ seconds and $16$ minutes $31$ for PBC and OBC, respectively. Circuit depth with respect to the Trotter steps for $H_{\mathrm{iso}}$ with $N=96, 100$ qubits for PBC and OBC, respectively, after Qiskit transpile with the optimization level 3.}
\label{tab:energy}
\end{table*}
\section{Case Studies on NISQ Algorithms}
\label{sec:nisq}
We instantiate the common and NISQ-specific terms of
Eq.~\eqref{eq:nisq} on two workloads of current scientific interest:
Hamiltonian time evolution and the variational quantum eigensolver.

\subsection{Time-Evolution Simulation of a Heisenberg Spin Chain}
\label{sec:te}
\paragraph{Setup.}
We benchmark the time evolution of the isotropic Heisenberg
Hamiltonian $H_\text{iso}$ on IBM Eagle r3 superconducting
processors, extending the experimental setup of \cite{chowdhury2024enhancing} to large-scale noisy quantum hardware and adding an end-to-end energy-consumption estimate. Two
configurations are evaluated: $N{=}100$ qubits with open boundary
conditions (OBC) on \texttt{ibm\_brisbane}, and $N{=}96$ qubits with periodic boundary conditions (PBC) on \texttt{ibm\_sherbrooke}. The
PBC count is reduced from 100 to 96 in order to permit a coupling between the first
and last qubits. 

\vspace{-10pt}
\paragraph{Quantum-error-mitigation stack.}
QEC is impractical at this scale on contemporary superconducting
hardware due to its qubit overhead, so we instead rely on QEM, which
accepts hardware imperfection at the cost of an additional circuit
duplication and sampling. Four QEM techniques are stacked:
ZNE with scaling factors $\{1,3,5\}$
(three folding copies per circuit); PT with $P=10$
randomized copies per fold; DD inserted into
idle windows; and M3 for
readout-error correction. Each circuit is measured at
$S = 10^5$ shots. Substituting these settings into
Eq.~\eqref{eq:nisq} gives
$E_\text{QEM} = (1{+}3{+}5)\cdot P\cdot S\cdot E_g\cdot N_\text{gates}$,
where $N_\text{gates}$ is the post-transpile gate count.

\vspace{-5pt}
\paragraph{Energy estimate.}
Following the per-technology energy-per-gate model
of~\cite{enriquez2023estimating}, we adopt $E_g \approx 0.18$\,J for
superconducting two-qubit gates (compared with $\sim 15$\,J for
trapped-ion gates), which implicitly absorbs operational and cooling overhead into the gate-energy term. The total experiment energy is obtained by counting all gates across the duplicated circuits of the QEM stack (Qiskit transpile, optimization level 3) and multiplying by the shot count $S =10^5$ shots as presented in Table~\ref{tab:energy}.



\subsection{Variational Quantum Eigensolver (VQE)}
\label{sec:vqe}
VQE estimates ground-state energies of a Hamiltonian $H = \sum_p c_p P_p$
(decomposed into Pauli strings $P_p$) by variationally minimizing
$\langle\psi(\boldsymbol{\theta})|H|\psi(\boldsymbol{\theta})\rangle$
over a parameterized ansatz $|\psi(\boldsymbol{\theta})\rangle$. Its energy footprint differs from time evolution due to the short Ansatz and the iterative algorithm.
\vspace{-5pt}
\paragraph{Energy model.} For a workload with $K$ optimizer iterations,
$M$ groups of Pauli strings to estimate per iteration (where one group corresponds to one circuit), $S$ shots per circuit, and an Ansatz of $G$ two-qubit gates, the gate cost is
\begin{equation}
E_\text{gate}^\text{VQE} \;=\; G\cdot M\cdot S\cdot K\cdot E_g.
\label{eq:Evqe-gate}
\end{equation}
QEM (e.g., ZNE+PT) multiplies this further by
$(\sum_k \alpha_k)\cdot P$ as in Eq.~\eqref{eq:nisq}. The classical
optimizer, Hamiltonian grouping, and parameter update contribute to
$E_\text{cls}$ and become non-trivial when $K$ is large or the optimizer is non-trivial.

\vspace{-5pt}
\paragraph{Comparison.} The two NISQ workloads show that VQE energy is
dominated by the optimizer-loop multiplier $K$, whereas time-evolution
energy is dominated by circuit depth and the per-fold ZNE
multiplier. Improvements that reduce $K$ such as better warm-starts~\cite{zou2025generative},
adaptive ansatze~\cite{grimsley2019adaptive} can reduce the energy consumption of VQE; whereas improvements that reduce QEM overhead such as better-chosen ZNE scale factors~\cite{cai2023quantum} can help with energy consumption for time evolution.

\section{Energies of FTQC: A Pipeline View}
\label{sec:ftqc}
The FTQC era inherits the common overhead of Eq.~\eqref{eq:total} and pays
$E^\text{exec}_\text{FTQC}$ from Eq.~\eqref{eq:ftqc} on the regime-specific
side. We analyze the end-to-end compilation pipeline that maps a fault-tolerant quantum algorithm down to physical qubit-cycles, and detail the energy consumption of each compilation stage below.


\vspace{-5pt}

\paragraph{Stage 1: Fault-tolerant quantum algorithm to logical quantum circuit.}
A fault-tolerant quantum algorithm (e.g., Shor, Grover, quantum chemistry, Hamiltonian
simulation) is synthesized to a logical circuit with a fault-tolerant gate set. These sets typically consist of a non-Clifford gate combined with the Clifford group, with the Clifford$+T$ set being a standard choice. This synthesis process can be implemented using existing tools such as AlphaTensor \cite{ruiz2025quantum}. The circuit includes 
$N_L$ logical qubits, $N_T$ $T$-gates, and $N_C$ Clifford gates. This stage sets
$N_T$ in Eq.~\eqref{eq:ftqc}.
\vspace{-5pt}
\paragraph{Stage 2: Encoding into physical qubits.} 
We utilize the well explored surface code for the compilation pipeline and employ lattice surgery for logical operations, as this approach is compatible with a wide range of quantum platforms, including limited 2D nearest-neighbor architectures for superconducting qubits and more flexible architectures of neutral atoms. Although transversal Clifford gates achieve $\Theta(1)$ time complexity, their implementation is contingent upon specific hardware capabilities, such as 3D inter-patch coupling or atom-style qubit transport \cite{chen2026transversal}.
The energy of a logical qubit cell $E_\text{cyc}$ is fixed by
the physical platform together with the chosen code distance $d$,
which is in turn picked to meet the target logical error rate
$p_L \approx (p/p_\text{th})^{(d+1)/2}$ over the workload's
spacetime volume. This stage sets $E_\text{cyc}$ in Eq.~\eqref{eq:ftqc}.
\vspace{-5pt}
\paragraph{Stage 3: Logical-circuit compilation.} The Clifford$+T$
program of Stage~1 must be scheduled into a concrete spacetime layout
of data patches, magic-state factories, ancilla bus channels, and
merge/split operations. This can be achieved by open-source lattice surgery compilers such as
TQEC~\cite{tqec}. The
compiled layout determines the spacetime volume, which directly
multiplies the energy cost of one logical qubit cell in Stage 2, and the end-to-end
energy estimates can inherit the quality of the
schedule. This stage sets $V_\text{ls}$ in Eq.~\eqref{eq:ftqc}.
\vspace{-5pt}
\paragraph{Stage 4: Magic-state distillation/cultivation.} $T$-gates are realized by consuming high-fidelity $|T\rangle$ states
prepared off-line. Conventional magic-state distillation
protocols achieve qubit cost $\Theta(d^2\!\cdot\!k)$ for $k$ levels of
distillation~\cite{litinski2019magic}. For example,
a single-level $15$-to-$1$ factory at distance
$d_f\!\approx\!15$ produces output states with error $\sim\!10^{-8}$
at roughly $810$ qubit-cycles per output
$T$~\cite{litinski2019magic}. More
recent magic-state cultivation
\cite{gidney2024magic} grows a $T$ state directly inside a
small surface-code patch and then expands the patch fault-tolerantly
to the target distance, reducing spacetime volume per $T$ by roughly
an order of magnitude relative to distillation at comparable output
fidelity. This stage sets $E_\text{ms}$ in Eq.~\eqref{eq:ftqc}.
\vspace{-5pt}
\paragraph{Stage 5: Real-time decoding.} Each surface-code cycle
produces $\Theta(d^2)$ syndrome bits per logical patch that must be
decoded inside the cycle latency. Reference implementations can place decoders in the cryostat or at room temperature. 
Each placement trades off accuracy, latency, energy, and thermal load on the dilution refrigerator.
Quantum system designers must carefully choose the decoder to meet all system requirements while minimizing the energy consumption.
Our hardware implementation of a BPOSD decoder~\cite{bposdcpp} indicates a median decoding latency of 25$ns$ for $d=11$, while a minimum weight perfect matching (MWPM) decoder~\cite{Micro_blossom} shows 33$ns$ with similar logical error rates at $4.6\times$ higher power.
Going to $d=32$, MWPM needs $12.2\times$ higher power.
The gap will grow exponentially larger if scaling the decoding system to more logical qubits, due to low scalability of MWPM~\cite{delfosse2021almost}. This
stage sets $E_\text{dec}$ in Eq.~\eqref{eq:ftqc}.
We show a collection of our implementations~\cite{lotterybp} in Table~\ref{tab:decoder_hw}.

\begin{table}[h]
    \centering
    \setlength{\tabcolsep}{4pt}
    \caption{Average metrics of BPOSD (B) and MWPM (M) hardware decoders for one logical qubit across distances.}
    \label{tab:decoder_hw}
    \begin{tabular}{c|c|c|c|c|c|c}
         \hline
        \multirow{2}{*}{$\boldsymbol{d}$} & \multicolumn{2}{c|}{\textbf{Area} ($mm^2$)} & \multicolumn{2}{c|}{\textbf{Power} ($W$)} & \multicolumn{2}{c}{\textbf{Latency} (ns)} \\
         \cline{2-7}
         & \textbf{B} & \textbf{M} & \textbf{B} & \textbf{M} & \textbf{B} & \textbf{M} \\
         \hline
         \hline
        7 & 0.90 & 0.38 & 0.27 & 0.19 & 19.6 & 14.4 \\
         \hline
        11 & 1.62 & 1.76 & 0.28 & 0.92 & 26.6 & 35.5 \\
         \hline
        13 & 4.35 & 3.10 & 0.36 & 1.62 & 32.8 & 49.6 \\
         \hline
        32 & 57.45 & 59.09 & 2.49 & 30.33 & 145.0 & 300.5 \\
         \hline
    \end{tabular}

\end{table}
\vspace{-5pt}
\paragraph{Stage 6: System maintenance at scale.} Cryogenic load grows
with physical-qubit count. A 1M-qubit superconducting machine implies
multiple parallel dilution refrigerators and tens to hundreds of
kilowatts of continuous power, dwarfing the per-job dynamic
energy~\cite{martin2022energy}. This stage feeds $E_\text{sys}$ in Eq.~\eqref{eq:Ecls},
not $E^\text{exec}_\text{FTQC}$.
\vspace{-5pt}
\paragraph{Where the joules go.} In logical quantum circuits, $E_\text{ms}\!\sim\!10^{3}\text{-}10^{4} E_\text{cyc}$ for
distillation, ${\sim}10\times$ less for
cultivation~\cite{litinski2019magic,gidney2024magic}. $T$-heavy
workloads (Shor, chemistry) are bound by $N_T E_\text{ms}$;
Clifford-heavy workloads by $V_\text{ls} E_\text{cyc}$; both pay
$E_\text{sys}$ as a continuous cryogenic tax that, at million-qubit
scale, runs at ${\sim}0.1$-$1$\,MW unconditionally~\cite{martin2022energy}.


\vspace{-5pt}
\paragraph{Implications.} The crossover question is whether classical
alternatives consume more energy than the FTQC version of the same
algorithm. Prior work~\cite{chen2023quantum} suggests that quantum
acceleration must exceed $\sim 100\times$ to offset cryogenic
overheads alone, since dilution refrigerators operating near 10\,mK
draw tens of kilowatts of wall-plug power regardless of the
computational workload they support. Eq.~\eqref{eq:ftqc} makes this
concrete: the inequality is a comparison of measurable quantities in
joules, not gate counts. This reframing has several consequences
worth drawing out. First, the dominant term tells the architect
where to push: lowering $E_\text{ms}$ through better magic-state
protocols pays off for $T$-heavy workloads, while shrinking
$V_\text{ls}$ through tighter scheduling and lower code distance
pays off for Clifford-heavy ones, and effort spent on the wrong
lever yields negligible returns. Second, $E_\text{dec}$ deserves
attention disproportionate to its apparent size in the equation:
decoding runs continuously on classical hardware throughout the
computation, and if it cannot keep pace with the syndrome extraction
rate, the logical clock stalls and the cryostat continues to draw
power without producing useful work. A slow decoder thus inflates
$E_\text{sys}$ multiplicatively rather than adding to the budget
linearly, which makes decoder throughput and energy efficiency a
leveraged investment, improvements there reduce both $E_\text{dec}$
itself and the idle tax it would otherwise impose. Third,
$E_\text{sys}$ behaves qualitatively differently from the
operation-based terms: it is paid in wall-clock time rather than in
gates, so any calibration pause, serial bottleneck, or
classical-feedback latency inflates the total energy without
producing computation. This favors algorithms that keep the device
saturated and penalizes those with sparse parallelism. Finally,
energy accounting closes a loophole that gate-count comparisons
leave open: a quantum algorithm with a favorable asymptotic speedup
can still lose on joules if its prefactor is dominated by
magic-state synthesis, if its decoder bottlenecks the logical clock,
or if its runtime forces the cryostat to absorb a large idle tax.
The practical suggestion is that algorithm designers report a joule
budget alongside gate counts, and that hardware roadmaps quote
$E_\text{cyc}$, decoder throughput per watt, and refrigerator
wall-plug efficiency as first-class figures of merit, since these
are the quantities that actually determine whether a given
application crosses the classical threshold.
\vspace{-5pt}

\section{Conclusion}
\label{sec:concl}
Given the rise of quantum utility, it is critical to understand its energy consumption, a critical metric that has become a bottleneck in the AI era.
We separated the energy sources of a quantum workload into those
common to NISQ and FTQC and those specific to each regime, instantiated
the NISQ side on 96- and 100-qubit Heisenberg time-evolution experiments
and a representative VQE chemistry workload, and traced the FTQC
pipeline from logical circuit through magic-state distillation/cultivation to
syndrome decoding. The case studies show that error-handling
overhead, not gate count, not cryogenics, sets the energy
budget today, and that the dominant cost will shift from sampling
multipliers in the NISQ regime to encoding and
magic state in the FTQC regime. Making these costs explicit is a
prerequisite for deciding when and where quantum computing
should be applied. For future works, we plan to demonstrate an end-to-end energy-consumption pipeline for fault-tolerant quantum algorithms, paving the road toward efficient quantum advantage.

\bibliographystyle{ACM-Reference-Format}
\bibliography{ref}

\end{document}